\documentclass[aps,prd,twocolumn,groupedaddress,amsmath,amssymb]{revtex4}
\usepackage{graphicx}
\usepackage{dcolumn}
\usepackage{bm}
\usepackage{verbatim}
\usepackage{color}

\begin{document}

\title{General formulas for conserved charges and black hole entropy in Chern-Simons-like theories of gravity}

\author{Mohammad Reza Setare}
\email[]{rezakord@ipm.ir}
\affiliation{{\small {\em  Department of Science, University of Kurdistan, Sanandaj, Iran.}}}
\author{Hamed Adami}
\email[]{hamed.adami@yahoo.com}
\affiliation{{\small{\em  Department of Science, University of Kurdistan, Sanandaj, Iran.}}}

\begin{abstract}
This paper deals with the problem of defining off-shell conserved charges in a set of theories known as Chern-Simons-like theories of gravity (CSLTG). The method is derived in a general way, which may find applications in a wide set of theories, and then specified to the case of  Generalized minimal massive gravity (GMMG), where known results of the central charge and black hole entropy are reproduced. The results for the charges are useful to the community and can be applied to all three-dimensional gravity theories within the class of CSLTG which includes almost all of them. The given specific examples show a consistency check of the more general method.
\end{abstract}

\maketitle

\section{Introduction}
 There is a class of gravitational theories in $(2+1)$-dimension (e.g. Topological massive gravity (TMG) \cite{23}, New massive gravity (NMG) \cite{24}, Minimal massive gravity (MMG) \cite{25}, Zewi-dreibein gravity (ZDG) \cite{26}, Generalized minimal massive gravity (GMMG) \cite{27}, etc), called the Chern-Simons-like theories of gravity (CSLTG) \cite{1}. In this work, we try to obtain formulas for conserved charges and black hole entropy in the context of CSLTG.\\
  Arnowitt, Deser, and Misner have introduced a formalism (ADM formalism) \cite{3}. Since space-time decomposition has been used in ADM formalism then, it is a non-covariant one. Conserved charges of an asymptotically flat spacetime solution of a general theory of relativity can be obtained by ADM formalism. ADM formalism has been extended to include asymptotically AdS spacetimes \cite{2}. Deser and Tekin have extended this approach to calculate conserved charges of asymptotically (A)dS solutions in higher curvature gravity models \cite{4}. Nowadays this method is referred to as ADT formalism which is covariant one. Another method to obtain conserved charges in gravity theories is covariant phase space method which has been proposed by introducing the concept of symplectic current in gravity theories \cite{20,54}. There are some non-covariant gravity theories, for instance TMG, which means that the variation of Lagrangian of those theories induced by diffeomorphism generator $\xi$ is not equal to its Lie derivative along vector field $\xi$. Tachikawa extended covariant phase space method to include non-covariant theories of gravity \cite{21}. ADT formalism and covariant phase space method are the cornerstones of our discussion. ADT formalism and covariant phase space method are not applicable to obtain the entropy of black holes in CSLTG because CSLTG is written in the first formalism and in this formalism we have the Lorentz gauge transformations as well as diffeomorphism. Regardless of the Lorentz gauge transformations, the conserved charge associated with the horizon-generating Killing vector field evaluated at the bifurcation surface vanishes. One way to deal with this problem is that one chooses the Cauchy surface such that the event horizon does not lie on the bifurcation surface \cite{53}. Another way is that one takes the Lorentz gauge transformations into account in addition to the diffeomorphism. In this way one can take the bifurcation surface as the integration surface. Here, we consider such a way to find a general formula for the black hole entropy in CSLTG. Ordinary ADT formalism is the on-shell one, but the generalization of that formalism presented in \cite{6,7,47} has been done in an off-shell way. One can use off-shell ADT formalism to find the conserved charges corresponding to the asymptotic symmetries in the Einstein-Maxwell theory (see \cite{48}, for instance). The formalism which will be presented here is off-shell and quasi-local. As we mentioned in above, the Chern-Simons-like theories of gravity are formulated in the first order formalism and the quasi-local method of obtaining conserved charges \cite{6,7,47} is formulated in the metric formalism, so we must provide the quasi-local method in the first order formalism.
\section{Chern-Simons-like theories of gravity}
Let $e^{a} = e^{a}_{\hspace{1.5 mm} \mu} dx^{\mu}$ be a Lorentz vector-valued 1-form, where $e^{a}_{\hspace{1.5 mm} \mu}$ denotes the dreibein \footnote{We use the lower case Greek letters for the spacetime indices, and the internal Lorentz indices are denoted by the lower case Latin letters. The metric signature is mostly plus.}. Spacetime metric $g_{\mu \nu}$ and dreibein are related as $g_{\mu \nu}=\eta_{ab} e^{a}_{\hspace{1.5 mm} \mu} e^{a}_{\hspace{1.5 mm} \nu}$, where $\eta_{ab}$ is Minkowski metric. It is clear that the spacetime metric is invariant under a Lorentz gauge transformations (or equivalently local Lorentz transformations, see Appendix J of \cite{50}) $e^{a} \rightarrow \Lambda ^{a}_{\hspace{1.5 mm} b} e^{b}$, where $\Lambda \in SO(2,1)$. Let $\mathcal{A}^{a \cdots}_{b \cdots}$ be a Lorentz-tensor-valued $p$-form. Exterior Lorentz covariant derivative (ELCD) of $\mathcal{A}^{a \cdots}_{b \cdots}$ can be defined as
\begin{equation}\label{110}
D(\omega)\mathcal{A}^{a \cdots}_{b \cdots}= d \mathcal{A}^{a \cdots}_{b \cdots} + \omega^{a}_{\hspace{1.5 mm} c} \mathcal{A}^{c \cdots}_{b \cdots}+ \cdots - \omega^{c}_{\hspace{1.5 mm} b} \mathcal{A}^{a \cdots}_{c \cdots} - \cdots,
\end{equation}
where $\omega^{a}_{\hspace{1.5 mm} b}= \omega^{a}_{\hspace{1.5 mm} b \mu} dx^{\mu}$ is spin-connection 1-form and we dropped wedge product for simplicity. It can be shown that ELCD is Lorentz covariant if spin-connection is transformed as
\begin{equation}\label{111}
\omega ^{a}_{\hspace{1.5 mm} b} \rightarrow \tilde{\omega}^{a}_{\hspace{1.5 mm} b}= \Lambda ^{a}_{\hspace{1.5 mm} c} \omega ^{c}_{\hspace{1.5 mm} d} (\Lambda^{-1}) ^{d}_{\hspace{1.5 mm} b}+ \Lambda ^{a}_{\hspace{1.5 mm} c} d (\Lambda^{-1})^{c}_{\hspace{1.5 mm} b},
\end{equation}
under the Lorentz gauge transformations. Albeit the spin-connection is a covariant quantity under diffeomorphism, it is not covariant under the Lorentz gauge transformation. In 3D, one can define the dualized spin-connection 1-form as $\omega^{a}=\frac{1}{2}\varepsilon^{abc} \omega_{bc}$. In this way, one can define dualized curvature and torsion 2-form as $R(\omega) = d \omega+\omega \times \omega$ and $T(\omega) = D(\omega)e=de+\omega \times e$, respectively. We use a 3D-vector algebra notation for Lorentz vectors in which contractions with $\eta _{ab}$ and $\varepsilon _{abc}$ are denoted by dots and crosses, respectively. One can formulate CSLTG in terms of $e^{a}$, $\omega^{a}$ and some Lorentz vector valued 1-form auxiliary fields. So, the Lagrangian 3-form of CSLTG is given by
\begin{equation}\label{1}
 L= \frac{1}{2} \textbf{g}_{rs} a^{r} \cdot d a^{s}+ \frac{1}{6} \textbf{f}_{rst} a^{r} \cdot a^{s} \times a^{t}.
\end{equation}
In the above Lagrangian $ a^{ra}=a^{ra}_{\hspace{3 mm} \mu} dx^{\mu} $ is the Lorentz vector valued 1-form, where $r=1, ..., N$ refers to the "flavor" index. Also, $\textbf{g}_{rs} $ is a symmetric constant metric on the flavor space and $\textbf{f}_{rst}$ is the totally symmetric "flavor tensors" interpreted as the coupling constant. It is worth saying that $ a^{ra}$ is a collection of the dreibein, the dualized spin-connection and the auxiliary fields, i.e. $a^{ra}=\{ e^{a},\omega ^{a}, h^{a}, f^{a} , \cdots \}$, where $h^{a}$ and $f^{a}$ are Lorentz vector valued auxiliary 1-form fields. Also, for all interesting CSLTG we have $\textbf{f}_{\omega rs} = \textbf{g}_{rs}$.
\section{Lorentz-Lie derivative and total variation}
Let $\pounds_{\xi}$ denote an ordinary Lie derivative along a vector field $\xi$. Ordinary Lie derivative of a Lorentz tensor-valued $p$-form is not covariant under the Lorentz gauge transformations. So, we need to modify the Lie derivative such that it becomes covariant under the Lorentz gauge transformations. Therefore we define the Lorentz-Lie derivative (LL-derivative) as
\begin{equation}\label{112}
\mathfrak{L}_{\xi} \mathcal{A}^{a \cdots}_{b \cdots}= \pounds_{\xi} \mathcal{A}^{a \cdots}_{b \cdots} + \lambda^{\hspace{1 mm} a}_{\xi \hspace{1 mm} c} \mathcal{A}^{c \cdots}_{b \cdots}+ \cdots - \lambda^{\hspace{1 mm} c}_{\xi \hspace{1 mm} b}\mathcal{A}^{a \cdots}_{c \cdots} - \cdots,
\end{equation}
where $\lambda^{\hspace{1 mm} a}_{\xi \hspace{1 mm} b}$ is the generator of the Lorentz gauge transformations (where $\Lambda=\exp \lambda$) and it is antisymmetric\cite{22}. $\lambda_{\xi}$ is transformed as a connection for the LL-derivative,
\begin{equation}\label{124}
\lambda ^{\hspace{1 mm} a}_{\xi \hspace{1 mm} b} \rightarrow \tilde{\lambda}^{\hspace{1 mm} a}_{\xi \hspace{1 mm} b}= \Lambda ^{a}_{\hspace{1.5 mm} c} \lambda ^{\hspace{1 mm} c}_{\xi \hspace{1 mm} d} (\Lambda^{-1})^{d}_{\hspace{1.5 mm} b} + \Lambda ^{a}_{\hspace{1.5 mm} c} \pounds_{\xi} (\Lambda^{-1})^{c}_{\hspace{1.5 mm} b}.
\end{equation}
The transformation law of $ \lambda ^{\hspace{1 mm} a}_{\xi \hspace{1 mm} b}$ under Lorentz gauge transformations can be simplified as follows:
  \begin{equation}\label{100}
    \begin{split}
       \tilde{\lambda}_{\xi}  &= \Lambda \lambda_{\xi}\Lambda^{-1} + \Lambda \pounds_{\xi} \Lambda^{-1} \\
         & =e^{\lambda_{\xi}} \lambda_{\xi} e^{-\lambda_{\xi}} + e^{\lambda_{\xi}} \pounds_{\xi} e^{-\lambda_{\xi}} \\
         & =\lambda_{\xi}-\pounds_{\xi} \lambda_{\xi},
    \end{split}
  \end{equation}
where we dropped Lorentz indices for simplicity. Therefore we do not need to consider behaviour of $\Lambda ^{\hspace{1 mm} a}_{\xi \hspace{1 mm} b}$ under Lorentz gauge transformations, in general.\\
Now, we introduce the total variation as combination of variations due to the diffeomorphism $\delta_{\text{D}}$ and the infinitesimal Lorentz transformation $\delta_{\text{L}}$, i.e. $\delta_{\xi} \equiv \delta_{\text{D}}+ \delta_{\text{L}}$ \cite{28}. Since $\mathcal{A}^{a \cdots}_{b \cdots}$ is transformed as $\mathcal{A}^{a \cdots}_{b \cdots} \rightarrow \Lambda^{a}_{\hspace{1.5 mm} c} \cdots (\Lambda^{-1})^{d}_{\hspace{1.5 mm} b} \cdots \mathcal{A}^{c \cdots}_{d \cdots}$ under Lorentz gauge transformation, the variation of $\mathcal{A}^{a \cdots}_{b \cdots}$ induced by $\lambda^{\hspace{1 mm} a}_{\xi \hspace{1 mm} b}$ is given by
\begin{equation}\label{113}
\begin{split}
\delta_{\text{L}} \mathcal{A}^{a \cdots}_{b \cdots}= & \tilde{\mathcal{A}}^{a \cdots}_{b \cdots} -\mathcal{A}^{a \cdots}_{b \cdots}  \\
= &  \lambda^{\hspace{1 mm} a}_{\xi \hspace{1 mm} c} \mathcal{A}^{c \cdots}_{b \cdots}+ \cdots - \lambda^{\hspace{1 mm} c}_{\xi \hspace{1 mm} b}\mathcal{A}^{a \cdots}_{c \cdots} - \cdots.
\end{split}
\end{equation}
Also, we have $\delta_{\text{D}}\mathcal{A}^{a \cdots}_{b \cdots} =  \pounds_{\xi} \mathcal{A}^{a \cdots}_{b \cdots}$. In this way, the total variation of a Lorentz tensor-valued $p$-form is equal to its LL-derivative, that is $\delta_{\xi} \mathcal{A}^{a \cdots}_{b \cdots} =\mathfrak{L}_{\xi} \mathcal{A}^{a \cdots}_{b \cdots}$. Spin-connection is a covariant quantity under diffeomorphism then $\delta_{\text{D}} \omega^{a}_{\hspace{1.5 mm}b}= \pounds_{\xi} \omega^{a}_{\hspace{1.5 mm}b} $ holds. In contrast, it is not covariant under Lorentz gauge transformations, see Eq.\eqref{111}. Then the variation of spin-connection induced by $\lambda_{\xi}$ is
\begin{equation}\label{114}
\begin{split}
\delta_{\text{L}} \omega^{a}_{\hspace{1.5 mm}b}= &  \tilde{\omega}^{a}_{\hspace{1.5 mm}b} -  \omega^{a}_{\hspace{1.5 mm}b}  \\
= & \lambda^{\hspace{1 mm} a}_{\xi \hspace{1 mm} c} \omega^{c}_{\hspace{1.5 mm}b} - \lambda^{\hspace{1 mm} c}_{\xi \hspace{1 mm} b} \omega^{a}_{\hspace{1.5 mm}c} - d \lambda^{\hspace{1 mm} a}_{\xi \hspace{1 mm} b} .
\end{split}
\end{equation}
Therefore total variation of dualized spin-connection becomes $\delta_{\xi} \omega^{a}=\mathfrak{L}_{\xi} \omega^{a}- d \chi_{\xi}^{a}$, where $\chi_{\xi}^{a}= \frac{1}{2} \varepsilon^{a}_{\hspace{1.5 mm} bc} \lambda^{ab}_{\xi}$. Here, what we introduced as total variation of the dualized spin-connection $\delta_{\xi} \omega^{a}$ is the same as LL-derivative in Ref.\cite{22}. We distinguish them because of two reasons. First, we have numerous dynamical fields in CSLTG and we need to have a unique definition for LL-derivative. Second, our definition for LL-derivative of dualized spin-connection explicitly shows that dualized spin-connection is not covariant under Lorentz gauge transformations. Hence total variation of $a^{ra}$ is given by
\begin{equation}\label{2}
  \delta_{\xi} a^{ra} = \mathfrak{L}_{\xi} a^{ra} - \delta^{r}_{\omega} d \chi_{\xi}^{a},
\end{equation}
where $\delta^{r}_{s}$ is the Kronecker delta. Total variation of $a^{ra}$ is covariant under the Lorentz gauge transformations as well as diffeomorphisms. Thus, we are allowed to use covariant phase space method to obtain conserved charges of solutions of CSLTG.
\section{Variation of Lagrangian}
The variation of the Lagrangian in Eq.\eqref{1} is
\begin{equation}\label{3}
  \delta L = \delta a^{r} \cdot E_{r} + d \Theta (a, \delta a),
\end{equation}
where
\begin{equation}\label{4}
   E_{r}^{\hspace{1.5 mm} a} = \textbf{g}_{rs} d a^{sa} + \frac{1}{2} \textbf{f}_{rst} (a^{s} \times a^{t})^{a}.
\end{equation}
 $ E_{r}^{\hspace{1.5 mm} a} =0$ is the equation of motion and
\begin{equation}\label{115}
\Theta (a, \delta a)=\frac{1}{2} \textbf{g}_{rs} \delta a^{r} \cdot a^{s}
\end{equation}
 is the surface term. The total variation of the Lagrangian due to diffeomorphism generator $\xi$ can be written as
\begin{equation}\label{6}
  \delta_{\xi} L = \mathfrak{L}_{\xi} L + d \psi_{\xi}=d \left( i_{\xi} L + \psi_{\xi} \right),
\end{equation}
with $\psi _{\xi} = \frac{1}{2} \textbf{g}_{\omega r} d \chi_{\xi} \cdot a^{r}$. The equation \eqref{6} is equivalent to the statement that a symmetry is a transformation which leaves the Lagrangian density invariant, up to a total
divergence. Despite the fact that a Lagrangian is not invariant under general coordinates transformations and/or general Lorentz gauge transformations, if the total variation of a given Lagrangian can be written as \eqref{6} then $\xi$ could be a symmetry generator. Although the Lagrangian 3-form \eqref{1} is not invariant under general Lorentz gauge transformations but it is invariant under the infinitesimal Lorentz gauge transformation. Also, it is enough in obtaining generally covariant equations of motion that Lagrangian behaves like \eqref{6} under total variation \cite{21}. Since the Lagrangian 3-form \eqref{1} is not invariant under general Lorentz gauge transformations then we expect that  the surface term is not a covariant quantity under general Lorentz gauge transformations. Hence, the total variation of the surface term differs from its Lorentz-Lie derivative and one can find that  the total variation of the surface term is given by
\begin{equation}\label{8}
  \delta_{\xi} \Theta (a, \delta a) = \mathfrak{L}_{\xi} \Theta (a, \delta a) + \Pi_{\xi},
\end{equation}
where $\Pi _{\xi}=\frac{1}{2} \textbf{g}_{\omega r} d \chi_{\xi} \cdot \delta a^{r}$. From Eq.\eqref{6} one can conclude that the Lagrangian 3-form has symmetry up to a total derivative. For some models $\psi_{\xi}$ vanishes and we refer to them as the Lorentz-diffeomorphism covariant theories because they are globally covariant under the Lorenz gauge transformations as well as diffeomorphism \cite{38}.
\section{Extended off-shell ADT current and Charge}
By considering that the variation in Eq.\eqref{3} is the total variation generated by $\xi$, \footnote{For generality, we assume that the vector field $\xi$ depends on the dynamical fields, which means that $\xi=\xi(a,x)$}, we find that
\begin{equation}\label{34}
  \begin{split}
     d J_{\xi}= & \left( i_{\xi} \omega -  \chi_{\xi}\right) \cdot \left( D E_{\omega}+ e \times E_{e} + a^{r^{\prime}} \times E_{r^{\prime}}\right) \\
       & +i_{\xi} a^{r^{\prime}} \cdot D E_{r^{\prime}}-i_{\xi}D a^{r^{\prime}} \cdot E_{r^{\prime}}\\
       & +i_{\xi} e \cdot D E_{e}-i_{\xi}T \cdot E_{e}-i_{\xi}R \cdot E_{\omega},
  \end{split}
\end{equation}
with
\begin{equation}\label{9}
  J_{\xi}= \Theta (a, \delta_{\xi} a) - i_{\xi} L - \psi_{\xi} + i_{\xi} a^{r} \cdot E_{r} - \chi _{\xi} \cdot E_{\omega},
\end{equation}
here $i_{\xi}$ denotes the interior product in $\xi$ and $r^{\prime}$ runs over all the flavor indices except $e$ and $\omega$. The authors of \cite{20} have deduced that the right hand side of Eq.\eqref{34} is a linear combination of the Bianchi identities, namely $D R=0$ and $D T = R \times e$. To clarify this argument, let us consider the action of CSLTG, $S= \int_{\mathcal{M}} L$. Variation of the action induced by diffeomorphism generator $\xi$ is $ \int_{\mathcal{M}}  \delta_{\xi} L = \int_{\mathcal{M}} \left[ \delta_{\xi} a^{r} \cdot E_{r} + d \Theta (a, \delta_{\xi} a)\right]$, which can be written as
\begin{equation}\label{101}
  \begin{split}
 \int_{\partial \mathcal{M}} J_{\xi}= & \\
      \int_{ \mathcal{M}} \bigl[& \left( i_{\xi} \omega -  \chi_{\xi}\right) \cdot ( D(\omega) E_{\omega}+ e \times E_{e} + a^{r^{\prime}} \times E_{r^{\prime}}) \\
       & +i_{\xi} a^{r^{\prime}} \cdot D(\omega) E_{r^{\prime}}-i_{\xi}D(\omega) a^{r^{\prime}} \cdot E_{r^{\prime}}\\
       & +i_{\xi} e \cdot D(\omega) E_{e}-i_{\xi}T(\omega) \cdot E_{e}-i_{\xi}R(\omega) \cdot E_{\omega} \bigr],
  \end{split}
\end{equation}
where $\partial \mathcal{M}$ is boundary of spacetime $\mathcal{M}$. By assuming that the dynamical field vanish on the boundary of spacetime, L.H.S of above equation will vanish, i.e. under this assumption we have
\begin{equation}\label{102}
  \begin{split}
     0= \int_{ \mathcal{M}} \bigl[&  \left( i_{\xi} \omega -  \chi_{\xi}\right) \cdot ( D(\omega) E_{\omega}+ e \times E_{e} + a^{r^{\prime}} \times E_{r^{\prime}}) \\
       & +i_{\xi} a^{r^{\prime}} \cdot D(\omega) E_{r^{\prime}}-i_{\xi}D(\omega) a^{r^{\prime}} \cdot E_{r^{\prime}}\\
       & +i_{\xi} e \cdot D(\omega) E_{e}-i_{\xi}T(\omega) \cdot E_{e}-i_{\xi}R(\omega) \cdot E_{\omega} \bigr].
  \end{split}
\end{equation}
Integrand on the R.H.S of above equation is proportional to diffeomorphism generator $\xi$ and it does not depend on the derivatives of $\xi$. If we demand that above equation hold for any $\xi$, the integrand must be a linear combination of Bianchi identities. By applying the Bianchi identities, we have $d J_{\xi} = 0$, which is held identically. Strictly speaking, $J_{\xi}$ is an off-shell conserved current. Since $J_{\xi}$ is a closed form, by virtue of the Poincar$\acute{\text{e}}$ lemma, it is an exact form. After straightforward calculations, one finds that $J_{\xi}=d K_{\xi},$ where $K_{\xi}$ is the off-shell Noether potential and it is given by
\begin{equation}\label{10}
  K_{\xi}= \frac{1}{2} \textbf{g}_{rs} i_{\xi} a^{r} \cdot a^{s} -\textbf{g}_{\omega s} \chi_{\xi} \cdot a^{s}.
\end{equation}
 In order to find the linearized off-shell conserved current we consider the variation of the Noether potential with respect to the dynamical fields. To this end, we take an arbitrary variation from Eq.\eqref{9}, then by using  Eq.\eqref{3}, Eq.\eqref{8} and the fact that $\chi_{\xi}$ is linear in $\xi$ ($\delta \chi_{\xi}= \chi_{\delta \xi}$), we will have
 \begin{equation}\label{50}
 \begin{split}
d \left[ \delta K_{\xi} - i_{\xi} \Theta (a,\delta a) \right]=& \Pi_{\xi}- \delta \psi_{\xi} + \psi_{\delta \xi}\\
      &+ \delta \Theta (a, \delta _{\xi} a) - \delta _{\xi} \Theta (a, \delta a) \\
      & - i_{\delta\xi} L + i_{\delta\xi} a^{r} \cdot E_{r} - \chi _{\delta \xi} \cdot E_{\omega} - \psi_{\delta \xi}\\
      & +\delta a^{r} \cdot i_{\xi} E_{r} + i_{\xi} a^{r} \cdot \delta E_{r} - \chi _{\xi} \cdot \delta E_{\omega}.
 \end{split}
 \end{equation}
One can use the explicit form of $\Pi_{\xi}$ and $ \psi_{\xi}$ to show that the first line on the right hand side of Eq.\eqref{50} vanishes. By using Eq.\eqref{9} (after replacing $\xi$ by $\delta \xi$ in it), we can write Eq.\eqref{50} as $\mathcal{J}_{\text{ADT}}=d \mathcal{Q}_{\text{ADT}}$, where
\begin{equation}\label{11}
\begin{split}
   \mathcal{J}_{\text{ADT}} (a , \delta a, \delta _{\xi} a)=  & \delta a^{r} \cdot i_{\xi} E_{r} + i_{\xi} a^{r} \cdot \delta E_{r} - \chi _{\xi} \cdot \delta E_{\omega} \\
     & + \delta \Theta (a, \delta _{\xi} a) - \delta _{\xi} \Theta (a, \delta a) \\
     & - \Theta (a, \delta _{\delta \xi} a),
\end{split}
\end{equation}
and
\begin{equation}\label{12}
\begin{split}
   \mathcal{Q}_{\text{ADT}} (a , \delta a;\xi) = & \delta K_{\xi}- K_{\delta \xi} - i_{\xi} \Theta (a,\delta a) \\
     =& \left( \textbf{g}_{rs} i_{\xi} a^{r} - \textbf{g} _{\omega s} \chi _{\xi} \right) \cdot \delta a^{s}.
\end{split}
\end{equation}
We refer to $\mathcal{J}_{\text{ADT}}$ and $\mathcal{Q}_{\text{ADT}}$ as extended off-shell ADT current and charge, respectively \cite{29}. If we assume that $\xi $ is a Killing vector field, for which $ \delta _{\xi} a^{r}=0$ \footnote{ One can show that $ \delta \Theta (a, \delta _{\xi} a) - \delta _{\xi} \Theta (a, \delta a)- \Theta (a, \delta _{\delta \xi} a)= \tilde{g} _{rs} \delta _{\xi} a^{r} \cdot \delta a^{s}$.}, and the equations of motion are satisfied, i.e. $E_{r}=0$, then extended off-shell ADT current will be reduce to $\mathcal{J}_{\text{ADT}}= i_{\xi} a^{r} \cdot \delta E_{r}$ which is similar to the on-shell ADT current \cite{2,4} in the considered formalism. So, Eq.\eqref{11} can be an appropriate extension of the on-shell ADT current to the off-shell case. It is clear from Eq.\eqref{11} that $d\mathcal{J}_{\text{ADT}}=0$ is held identically. Hence, the current in Eq.\eqref{11} is conserved off-shell, i.e. $\mathcal{J}_{\text{ADT}}$ is closed and exact. Therefore, $\mathcal{Q}_{\text{ADT}} (a , \delta a;\xi)$ in Eq.\eqref{12} denots the extended off-shell ADT charge.
\section{Off-shell extension of the covariant phase space method}
We know that the variation of the Lagrangian 3-form in Eq.\eqref{1} is given by Eq.\eqref{3}. Let us take another variation from Eq.\eqref{3}. Regarding that two variations do not commute, i.e. $\delta _{1} \delta _{2} \neq \delta _{2} \delta _{1} $, we will have
\begin{equation}\label{13}
  d \Omega_{\text{LW}}(a; \delta_{1} a , \delta_{2} a)= \delta_{1} a^{r} \cdot \delta_{2} E_{r}-\delta_{2} a^{r} \cdot \delta_{1} E_{r},
\end{equation}
where $\Omega_{\text{LW}}$ is the Lee-Wald symplectic current which is defined as
\begin{equation}\label{14}
\begin{split}
   \Omega_{\text{LW}}(a; \delta_{1} a , \delta_{2} a)= & \delta_{1} \Theta (a, \delta _{2} a) - \delta _{2} \Theta (a, \delta_{1} a) \\
     & - \Theta (a, \delta _{[1,2]} a),
\end{split}
\end{equation}
with $\delta _{[1,2]}= \delta _{1} \delta _{2} - \delta _{2} \delta _{1}$ \cite{52}. It is clear from Eq.\eqref{13} that the symplectic current is conserved when $a^{r}$ and $\delta a^{r}$ satisfy the equations of motion and the linearized equations of motion, respectively. In CSLTG, symplectic current can be simplified as
\begin{equation}\label{15}
   \Omega_{\text{LW}}(a; \delta_{1} a , \delta_{2} a)= \textbf{g} _{rs} \delta _{2} a^{r} \cdot \delta_{1} a^{s},
\end{equation}
which clearly is closed,
\begin{equation}\label{116}
\begin{split}
& \delta _{1} \Omega_{\text{LW}}(a; \delta_{2} a , \delta_{3} a) + \Omega_{\text{LW}}(a; \delta_{1} a , \delta_{_{[2,3]}} a) \\
+&  \delta _{3} \Omega_{\text{LW}}(a; \delta_{1} a , \delta_{2} a) + \Omega_{\text{LW}}(a; \delta_{3} a , \delta_{_{[1,2]}} a) \\
+& \delta _{2} \Omega_{\text{LW}}(a; \delta_{3} a , \delta_{1} a) + \Omega_{\text{LW}}(a; \delta_{2} a , \delta_{_{[3,1]}} a)=0,
\end{split}
\end{equation}
and skew-symmetric,
\begin{equation}\label{117}
 \Omega_{\text{LW}}(a; \delta_{2} a , \delta_{1} a) = -  \Omega_{\text{LW}}(a; \delta_{1} a , \delta_{2} a).
\end{equation}
Also, $ \Omega_{\text{LW}}$ is non-degenerate, i.e.
\begin{equation}\label{118}
 \Omega_{\text{LW}}(a; \delta a , \hspace{0.3 cm})=0 \hspace{0.3 cm}  \text{if and only if}  \hspace{0.3 cm}  \delta a =0 .
\end{equation}
By setting $\delta_{1}= \delta$ and $\delta_{2}= \delta_{\xi}$ and by considering action of $(\delta \delta_{\xi} - \delta_{\xi} \delta)$ on $a^{r}$, we have
\begin{equation}\label{103}
    \begin{split}
       (\delta \delta_{\xi} - \delta_{\xi} \delta) a^{r} & = (\mathfrak{L}_{\xi} \delta a^{r}+\mathfrak{L}_{\delta\xi} a^{r} - \delta^{r}_{\omega} d \chi_{\delta\xi})- (\mathfrak{L}_{\xi} \delta a^{r}) \\
         & = \delta_{\delta_{\xi}}a^{r}
    \end{split}
  \end{equation}
where we used the fact that difference of two (dualized) spin-connection is a Lorentz (vector-valued) 2-tensor-valued 1-form, same argument is held for metric connection. In other words, we have $\delta _{[1,2]}= \delta _{\delta \xi}$. It is easy to see that the commutator of two variations will vanish when $\xi$ does not depend on the dynamical field. Therefore, it seems reasonable to assume that $ \delta _{1} \delta _{2} \neq \delta _{2} \delta _{1}$. In this case, one can show that, the right hand side of Eq.\eqref{13} can be written as a total derivative, namely
\begin{equation}\label{119}
  \Omega_{\text{LW}}(a; \delta a , \delta_{\xi} a)=  - d \{ \delta a^{r} \cdot i_{\xi} E_{r} + i_{\xi} a^{r} \cdot \delta E_{r} - \chi _{\xi} \cdot \delta E_{\omega}  \},
\end{equation}
and then one can show that Eq.\eqref{13} becomes $d\mathcal{J}_{\text{ADT}}=0$. In other words, the off-shell extension of the symplectic current $\hat{ \Omega}_{\text{LW}}$, which is defined by
\begin{equation}\label{120}
d \hat{ \Omega}_{\text{LW}} =  \Omega_{\text{LW}} +d \{ \delta a^{r} \cdot i_{\xi} E_{r} + i_{\xi} a^{r} \cdot \delta E_{r} - \chi _{\xi} \cdot \delta E_{\omega}\} ,
\end{equation}
and the extended off-shell ADT current are related as $ \hat{ \Omega}_{\text{LW}}= \mathcal{J}_{\text{ADT}}+ d Z_{\xi}$, where $Z_{\xi}$ is an arbitrary 1-form which is locally constructed out of $a^{r}$ and $\delta a^{r}$. Thus, the obtained conserved charges in the off-shell covariant phase space method and in the extended off-shell ADT method differ from each other by a constant. If we choose $Z_{\xi}=0$ then the obtained conserved charges in two different ways will be the same.
\section{Quasi-local conserved charges}
Let $ \mathcal{C}$ denote a Cauchy surface in spacetime and $\mathcal{V} \subseteq \mathcal{C}$ be a portion of the Cauchy surface $ \mathcal{C}$. Since $\mathcal{J}_{\text{ADT}}$ is identically conserved, that is $d\mathcal{J}_{\text{ADT}}=0$, then we can define perturbed quasi-local conserved charge associated with a vector field $\xi$ as
\begin{equation}\label{121}
\begin{split}
  \delta Q (\xi) =& -\frac{1}{8 \pi G}\int_{\mathcal{V}}\mathcal{J} _{ADT}(a, \delta a; \xi) \\
=& -\frac{1}{8 \pi G}\int_{\Sigma}\mathcal{Q} _{ADT} (a, \delta a; \xi),
\end{split}
\end{equation}
where $\Sigma$ denotes boundary of $ \mathcal{V}$ and it is a space-like codimension-2 surface. To find quasi-local conserved charges, we take an integration from Eq.\eqref{121} over one-parameter path on the solution space. To this end, suppose that $a^{r} (\mathcal{N})$ are collection of fields which solve the equations of motion of the CSLTG, where $\mathcal{N}$ is a free parameter in the solution space of equations of motion. Now, we replace $\mathcal{N}$ by $s\mathcal{N}$, where $0  \leq s  \leq 1$ is just a parameter. By expanding $a^{r} ( s \mathcal{N})$ in terms of $s$ we have $ a^{r} (s\mathcal{N})= a^{r} (0)+\frac{\partial}{\partial s}a^{r} (0) s + \cdots $. By substituting $ a^{r} = a^{r}(s\mathcal{N}) $ and $\delta a^{r} = \frac{\partial}{\partial s}a^{r} (0)$ into \eqref{121}, we can define the quasi-local conserved charge associated to the Killing vector field $\xi$, then we will have
\begin{equation}\label{17}
 Q ( \xi )= -\frac{1}{8 \pi G} \int_{0}^{1} ds \int_{\Sigma} \mathcal{Q}_{\text{ADT}} (a |s;\xi),
\end{equation}
where integration over $s$ denotes integration over the one-parameter path on the solution space. Due to the definition in Eq.\eqref{11}, the quasi-local conserved charge Eq.\eqref{17} is not only conserved for the Killing vectors which are admitted by spacetime everywhere, but also it is conserved for the asymptotic Killing vectors. In Eq.\eqref{17}, $s=0$ and $s=1$ may be taken just as a background solution and a given solution, respectively. In this way, background contribution is subtracted and then we will find a finite amount for charge.
 Eq.\eqref{17} gives us the quasi-local conserved charges. It is local in the mean that we can define charge perturbation \eqref{121} for arbitrary $\mathcal{V}$, because $d\mathcal{J}_{\text{ADT}}=0$  is held identically (off-shell). It is quasi-local, because Eq.\eqref{17} gives us the global conserved charges of a given spacetime independent of the choice of $\mathcal{V}$.
\section{Black hole entropy in CSLTG}
We use the Wald prescription to find a formula for black hole entropy in CSLTG. According to the Wald prescription, the black hole entropy is conserved charge associated with the horizon-generating Killing vector field $\zeta$. The horizon-generating Killing vector field $\zeta$ vanishes on the bifurcation surface $\mathcal{B}$. Now, let us take $\Sigma$ in Eq.\eqref{17} to be the bifurcation surface $\mathcal{B}$ then we will have
\begin{equation}\label{18}
 Q ( \zeta )  = \frac{1}{8 \pi G} \textbf{g} _{\omega r} \int_{\mathcal{B}}  \chi _{\zeta} \cdot a^{r}.
\end{equation}
So far, we have considered $\lambda_{\xi} ^{ab}$ to be a function of space-time coordinates and of the diffeomorphism generator $\xi$ \footnote{In general, we assume that $\lambda_{\xi}$  is a function of spacetime coordinates and of the diffeomorphism generator $\xi$. Also, we assume that it is linear in $\xi$. We found expressions for conserved charge. To use that expression we have to find an expression for $\lambda_{\xi}$ in terms of dynamical fields (i.e. we need a gauge fixing) when we calculate conserved charges of a solution. To this end, we demanded that the LL-derivative of $e^{a}$ explicitly vanishes when $\xi$ is a Killing vector field.}; and it is antisymmetric with respect to $a$ and $b$. To obtain an explicit expression for $\lambda_{\xi} ^{ab}$, in an appropriate manner, the authors in \cite{22} had to choose $\lambda_{\xi} ^{ab}$ in a way that the LL-derivative of $e^{a}$ vanishes when $\xi$ is a Killing vector field. They showed that $\lambda_{\xi} ^{ab}$ is given as $ \lambda_{\xi} ^{ab} = e^{\sigma [a} \pounds _{\xi} e^{b]}_{\hspace{1.5 mm} \sigma}$. Since $\chi_{\xi}^{a}= \frac{1}{2} \varepsilon^{a}_{\hspace{1.5 mm} bc} \lambda^{ab}_{\xi}$, using $\lambda_{\xi} ^{ab}$ one can show that \cite{28}:
\begin{equation}\label{19}
  \chi _{\xi} ^{a} = i_{\xi} \omega ^{a} + \frac{1}{2} \varepsilon ^{a}_{\hspace{1.5 mm} bc} e^{\nu b} (i_{\xi} T^{c})_{\nu} + \frac{1}{2} \varepsilon ^{a}_{\hspace{1.5 mm} bc} e^{b \mu} e^{c \nu} \nabla _{\mu} \xi _{\nu}.
\end{equation}
In this way, the local Lorentz gauge has been fixed. Since on the bifurcation surface $\zeta |_{\mathcal{B}}=0$ and $\nabla _{\mu} \zeta_{\nu} |_{\mathcal{B}}= \kappa n_{\mu \nu}$,  where $\kappa$ and $n_{\mu \nu}$ are the surface gravity and the bi-normal to the bifurcation surface respectively, we will have $ \chi _{\zeta}^{a}|_{\mathcal{B}}= \kappa N^{a}$ with $N^{a}=\frac{1}{2} \varepsilon^{abc}n_{bc}$. Therefore, using Eq.\eqref{18} the black hole entropy in the CSLTG  can be defined as
\begin{equation}\label{122}
  S= \frac{2\pi}{\kappa} Q ( \zeta ) =- \frac{1}{4 G} \textbf{g}_{\omega r} \int_{\mathcal{B}} N \cdot a^{r},
\end{equation}
The non-zero components of bi-normal to the bifurcation surface of a stationary black hole are $n_{tr}=-n_{rt}$, where $t$ and $r$ are respectively time and radial coordinates, and it is normalized to $-2$. Thus $N^{a}$ is normalized to $+1$, and then the only non-zero component of $N^{a}$ is $N^{\phi}=(g_{\phi \phi})^{-1/2}$, where $\phi$ is angular coordinate. Therefore, Eq.\eqref{122} can be written as
\begin{equation}\label{21}
  S= - \frac{1}{4G} \int_{\mathcal{B}} \frac{d \phi}{\sqrt{g_{\phi \phi}}} \textbf{g} _{\omega r} a^{r}_{\phi \phi},
\end{equation}
where $g_{\phi \phi}$ denotes the $ \phi-\phi $ component of the spacetime metric $g_{\mu \nu}$. One can use the above generic entropy formula to obtain the entropy of black hole solution of CSLTG.
\section{Two examples of CSLTG}
We consider two examples of CSLTG, one of them is the Einstein gravity with negative cosmological constant (EG) and the other one which is constructed out of $e$, $\omega$ and two auxiliary 1-form fields $h^{a}$ and $f^{a}$.
\subsection{Example(1): EG}
In this theory, we have $a^{r}=\{e,\omega \}$ and the non-zero components of the flavor metric and tensor are $\textbf{g}_{e \omega}=-1$ and $\textbf{f}_{eee}=-l^{-2}$ respectively, where $l$ is the radius of AdS$_{3}$. In this case, equations of motion \eqref{4} reduce to
\begin{equation}\label{35}
R(\Omega)+\frac{1}{2l^{2}} e \times e=0, \hspace{0.7 cm} T(\Omega)=0,
\end{equation}
where $\omega=\Omega$ is torsion-free spin-connection. The equation $ T(\Omega)=0$ implies
\begin{equation}\label{123}
  \Omega^{a}= \frac{1}{2}e^{a}_{\hspace{1.5 mm} \alpha}\epsilon^{\alpha}_{\hspace{1.5 mm} \nu \beta}  e^{\beta}_{\hspace{1.5 mm} c} \dot{\nabla}_{\mu} e^{c\nu} d x^{\mu}
\end{equation}
where $\dot{\nabla}$ denotes covariant derivative with respect to the Christoffel symbols.
\subsection{Example(2): GMMG}
In GMMG, there are four flavours of 1-form, $a^{r}= \{ e, \omega , h, f \}$ and the non-zero components of the flavour metric and the flavour tensor are
\begin{equation}\label{22}
\begin{split}
     & \textbf{g}_{e \omega}=\textbf{f}_{e \omega \omega}=-\sigma, \hspace{0.4 cm} \textbf{g}_{\omega f}=\textbf{f}_{f \omega \omega}=-\frac{1}{m^{2}}, \\
     & \textbf{g}_{e h}=\textbf{f}_{e h \omega}=1, \hspace{0.4 cm} \textbf{g}_{\omega \omega}=\textbf{f}_{\omega \omega \omega}=\frac{1}{\mu},\\
     & \textbf{f} _{eff}= -\frac{1}{m^{2}}, \hspace{0.4 cm} \textbf{f}_{eee}=\Lambda_{0},\hspace{0.4 cm} \textbf{f}_{ehh}= \alpha,
\end{split}
\end{equation}
where $\sigma$, $\Lambda _{0}$, $\mu$, $m$ and $\alpha$ are a sign, cosmological parameter with dimension of mass squared, mass parameter of the Lorentz Chern-Simons term, mass parameter of New massive gravity term and a dimensionless parameter, respectively. By substituting Eq.\eqref{22} into the equations \eqref{4}, we find equations of motion of the GMMG as
\begin{equation}\label{2.29}
   E _{e}=- \sigma R (\omega) + \frac{\Lambda _{0}}{2} e \times e + D(\omega) h - \frac{1}{2 m^{2}} f \times f  + \frac{\alpha}{2} h \times h =0 ,
\end{equation}
\begin{equation}\label{2.30}
  E _{\omega}=- \sigma T(\omega) + \frac{1}{\mu} R(\omega) - \frac{1}{m^{2}} D (\omega) f + e \times h =0,
\end{equation}
\begin{equation}\label{2.31}
   E _{f}=- \frac{1}{m^{2}} \left( R(\omega) + e \times f \right)=0 ,
\end{equation}
\begin{equation}\label{2.32}
  E _{h}=T(\omega) + \alpha e \times h =0.
\end{equation}
The equations \eqref{2.29}-\eqref{2.32} can be solved by the solutions of EG satisfying Eq.\eqref{35} with the following ansatz:
\begin{equation}\label{37}
   \omega^{a}= \Omega^{a}-\alpha H e^{a},\hspace{0.3 cm} h^{a}= H e^{a}, \hspace{0.3 cm} f^{a}= F e^{a},
\end{equation}
where $F$ and $H$ are constant parameters. In this way, equations of motion will be reduced to the following algebraic equations
among constant parameters
\begin{equation}\label{38}
  \begin{split}
       & \frac{\sigma}{ l^{2}} - \alpha (1 + \sigma \alpha ) H ^{2} + \Lambda _{0} - \frac{F^{2}}{ m^{2}}=0, \\
       & - \frac{1}{\mu l^{2}} + 2 (1 + \sigma \alpha ) H + \frac{2 \alpha}{m^{2}} F H + \frac{\alpha ^{2}}{\mu} H^{2}=0, \\
       & - F + \mu (1 + \sigma \alpha ) H + \frac{\mu \alpha}{m^{2}} FH=0.
  \end{split}
\end{equation}
Thus, all the solutions of EG are also the solutions of GMMG.
\section{Asymptotically AdS$_{3}$ spacetimes}
We consider AdS$_{3}$ spacetime as a background solution ($s=0$) and a given solution spacetime ($s=1$) which obeys the Brown-Henneaux boundary conditions. The Brown-Henneaux boundary conditions \cite{31} are appropriate for both EG and GMMG models. The dreibeins correspond to AdS$_{3}$ spacetime are
\begin{equation}\label{41}
   \bar{e}^{0} = \frac{r}{l} dt, \hspace{0.5 cm} \bar{e}^{1} = \frac{l}{r} dr, \hspace{0.5 cm} \bar{e}^{2} = r d \phi,
\end{equation}
where the bar sign refers to the background. The asymptotic Killing vectors preserving the Brown-Henneaux boundary conditions are given by
\begin{equation}\label{42}
\begin{split}
  \xi _{n} ^{\pm} = \frac{1}{2} e^{inx^{\pm}} \biggl[ & l \left( 1-\frac{l^{2} n^{2}}{2r^{2}} \right) \partial _{t} \\
     & -i n r \partial _{r} \pm \left( 1+\frac{l^{2} n^{2}}{2r^{2}} \right) \partial_{\phi} \biggr],
\end{split}
\end{equation}
where $x^{\pm}= t/l \pm \phi$ and $n \in \mathbb{Z}$. The asymptotic Killing vector fields $\xi _{n} ^{\pm}$ satisfy the Witt algebra, $i[\xi _{m} ^{\pm},\xi _{n} ^{\pm}]= (m-n) \xi _{m+n} ^{\pm}$. By Substituting equations \eqref{37}, Eq.\eqref{41} and \eqref{42} into Eq.\eqref{19}, we find that
\begin{equation}\label{7.9}
  i_{\xi _{n} ^{\pm}} \bar{\Omega} ^{a} - \bar{\chi} _{\xi _{n} ^{\pm}} ^{a} = \pm \frac{1}{l} \bar{e}^{a} _{\hspace{1.5 mm} \mu}(\xi _{n} ^{\pm})^{\mu} .
\end{equation}
Since the AdS$_{3}$ spacetime satisfy EG equations of motion \eqref{35} then one can show that
\begin{equation}\label{7.10}
  \begin{split}
       & \delta _{\xi _{n} ^{\pm}} \Omega ^{0} _{\hspace{1.5 mm} \phi} \pm \frac{1}{l} \delta _{\xi _{n} ^{\pm}} e^{0} _{\hspace{1.5 mm} \phi} = - \frac{i l n^{3}}{2r} e^{inx^{\pm}} , \\
       & \delta _{\xi _{n} ^{\pm}} \Omega ^{1} _{\hspace{1.5 mm} \phi} \pm \frac{1}{l} \delta _{\xi _{n} ^{\pm}} e^{1} _{\hspace{1.5 mm} \phi} = 0 , \\
       & \delta _{\xi _{n} ^{\pm}} \Omega ^{2} _{\hspace{1.5 mm} \phi} \pm \frac{1}{l} \delta _{\xi _{n} ^{\pm}} e^{2} _{\hspace{1.5 mm} \phi} = \pm \frac{i l n^{3}}{2r} e^{inx^{\pm}}.
  \end{split}
\end{equation}
The non-zero components of the flavour metric are given by Eq.\eqref{22}. Therefore, Eq.\eqref{17} will reduce to
\begin{equation}\label{7.11}
    Q ( \xi _{m} ^{\pm} )  =  \frac{c_{\pm}}{12 \pi l} \int_{\infty} (\xi _{m} ^{\pm}) \cdot \left( \Delta \Omega  _{\phi} \pm \frac{1}{l} \Delta e  _{ \phi} \right) d \phi ,
\end{equation}
with
\begin{equation}\label{40}
  c_{\pm}= \frac{3l}{2G} \left( \sigma + \frac{\alpha H}{\mu}+ \frac{F}{m^{2}} \mp \frac{1}{\mu l}\right),
\end{equation}
for asymptotic Killing vectors \eqref{42}, where Eq.\eqref{37} and Eq.\eqref{38} were used. In Eq.\eqref{7.11} the integration runs over a circle with radius of infinity. The above expression for the conserved charges associated with the asymptotic Killing vectors $\xi^{\pm}_{n}$ can be written as
\begin{equation}\label{23}
  Q(\xi^{\pm}_{n})=\frac{c_{\pm}}{12 \pi l} \int_{\infty} \xi^{\pm}_{n} \cdot A^{\pm}_{\phi} d\phi,
\end{equation}
where $A^{\pm a}= \Omega ^{a} \pm e^{a}/l$ are connections corresponding to two $SO(2, 1)$ gauge groups. In a similar way, Eq.\eqref{121} in the GMMG reduces to
\begin{equation}\label{7.15}
    \delta _{\xi _{n} ^{\pm}} Q ( \xi _{m} ^{\pm} )  = \frac{c_{\pm}}{12 \pi l} \lim _{r \rightarrow \infty} \int_{0}^{2 \pi} (\xi _{m} ^{\pm}) \cdot \delta _{\xi _{n} ^{\pm}} A^{\pm} _{\hspace{1.5 mm} \phi} d \phi.
\end{equation}
where we were set $\delta= \delta _{\xi _{n}^{\pm}}$. For  Ba\~nados,  Teitelboim and  Zanelli (BTZ) black hole spacetime \cite{60}
\begin{equation}\label{2.4.2}
  \begin{split}
       & e^{0}=\left( \frac{(r^{2}-r_{+}^{2})(r^{2}-r_{-}^{2})}{l^{2}r^{2}} \right)^{\frac{1}{2}}dt \\
       & e^{1}= r \left( d \phi -\frac{r_{+}r_{-}}{lr^{2}} dt \right) \\
       & e^{2}=\left( \frac{l^{2}r^{2}}{(r^{2}-r_{+}^{2})(r^{2}-r_{-}^{2})} \right)^{\frac{1}{2}}dr,
  \end{split}
\end{equation}
where $r_{\pm}$ are outer/inner horizon radiuses, at spatial infinity we have
\begin{equation}\label{7.13}
\begin{split}
    \Delta e^{a} _{\hspace{1.5 mm} \phi}& = 0, \hspace{0.7 cm} \Delta \Omega ^{0} _{\hspace{1.5 mm} \phi} =-\frac{r_{+}^{2}+r_{-}^{2}}{2 l r}\\
 \Delta \Omega ^{1} _{\hspace{1.5 mm} \phi}& = 0, \hspace {0.7 cm} \Delta \Omega ^{2} _{\hspace{1.5 mm} \phi} =-\frac{r_{+} r_{-}}{ l r},
\end{split}
\end{equation}
By substituting \eqref{42} and \eqref{7.13} into \eqref{23}, we find that
\begin{equation}\label{7.14}
    Q ( \xi _{m} ^{\pm} )  = \frac{c_{\pm}}{24} \left( \frac{r_{+} \mp r_{-}}{ l } \right) ^{2} \delta _{m,0} .
\end{equation}
Also, using Eq.\eqref{7.10} and \eqref{42}, one can show that Eq.\eqref{7.15} will reduce to
\begin{equation}\label{7.16}
    \delta _{\xi _{n} ^{\pm}} Q ( \xi _{m} ^{\pm} )  =  \frac{i n^{3} c_{\pm}}{12 } \delta _{m+n,0} .
\end{equation}
The algebra of conserved charges can be written as \cite{31,57}
\begin{equation}\label{5.6}
  \left\{ Q(\xi _{1}) , Q(\xi _{2}) \right\}_{\text{D.B.}} = Q \left(  \left[ \xi _{1} , \xi _{2} \right] \right) + \mathcal{C} \left( \xi _{1} , \xi _{2} \right)
\end{equation}
where $\mathcal{C} \left( \xi _{1} , \xi _{2} \right)$ is central extension term. Also, the left hand side of the equation \eqref{5.6} is Dirac brackets and it can be defined by
\begin{equation}\label{5.7}
  \left\{ Q(\xi _{1}) , Q(\xi _{2}) \right\}_{\text{D.B.}}= \frac{1}{2} \left( \delta _{\xi _{2}} Q(\xi _{1}) - \delta _{\xi _{1}} Q(\xi _{2}) \right).
\end{equation}
Therefore, by using the following formula, one can obtain the central extension term
\begin{equation}\label{5.8}
  \mathcal{C} \left( \xi _{1} , \xi _{2} \right)= \frac{1}{2} \left( \delta _{\xi _{2}} Q(\xi _{1}) - \delta _{\xi _{1}} Q(\xi _{2}) \right) - Q \left(  \left[ \xi _{1} , \xi _{2} \right] \right).
\end{equation}
In Eq.\eqref{5.6}, $\left[ \xi _{1} , \xi _{2} \right]$ is a modified version of the Lie brackets which is defined by \cite{58}
\begin{equation}\label{5.9}
  \left[ \xi_{1} , \xi_{2} \right] =  \pounds _{\xi_{1}} \xi_{2} - \delta ^{(g)}_{\xi _{1}} \xi_{2} + \delta ^{(g)}_{\xi _{2}} \xi_{1},
\end{equation}
where $\delta ^{(g)}_{\xi _{1}} \xi_{2}$ denotes the change induced in $\xi_{2}$ due to the variation of metric $\delta _{_{\xi _{1}}} g_{\mu\nu} = \pounds _{\xi_{1}} g_{\mu\nu}$. It is clear that Eq.\eqref{5.6} will reduce to ordinary Lie brackets $\left[ \xi_{1} , \xi_{2} \right]_{\text{Lie}}= \pounds _{\xi_{1}} \xi_{2}$ when $\xi$ does not depend on dynamical fields, i.e. $\delta ^{(g)}_{\xi _{1}} \xi_{2}= \delta ^{(g)}_{\xi _{2}} \xi_{1}=0$. By substitute \eqref{7.14} and \eqref{7.16} into \eqref{5.8}, we find the following expression for the central extension term
\begin{equation}\label{7.17}
  C _{E} (\xi _{m} ^{\pm} , \xi _{n} ^{\pm}) =- i \frac{c_{\pm}}{12} \left[ m^{3} - \left( \frac{r_{+} \mp r_{-}}{ l } \right) ^{2} m  \right] \delta _{m+n,0} .
\end{equation}
To obtain the usual $m$ dependence, that is $m (m ^{2} -1)$, it is sufficient one make a shift on $Q$ by a constant \cite{59}. Now, we set $Q (\xi _{n} ^{\pm}) \equiv \hat{L}^{\pm} _{n}$ and replace the Dirac brackets by commutators, namely $ \{ Q ( \xi _{m} ^{\pm} ) , Q ( \xi _{n} ^{\pm} ) \}_{\text{D.B.}} \equiv i [ \hat{L}^{\pm} _{m} , \hat{L}^{\pm} _{n}]$, then Eq.\eqref{5.6} becomes
\begin{equation}\label{7.18}
  [\hat{L}^{\pm} _{m} , \hat{L}^{\pm} _{n}] =(m-n) \hat{L}^{\pm} _{m+n} + \frac{c_{\pm}}{12} m (m ^{2} -1) \delta _{m+n,0} ,
\end{equation}
where $c_{\pm}$ are the central charges, and $\hat{L}^{\pm} _{n}$ are generators of Virasoro algebra. Therefore, the algebra among the asymptotic conserved charges, in asymptotically AdS$_{3}$ spacetimes, is isomorphic to two copies of the Virasoro algebra with central charges $c_{\pm}$ \cite{30}.\\
We can read off the eigenvalues of the Virasoro generators $\hat{L}^{\pm} _{n}$ from \eqref{7.14} as
\begin{equation}\label{7.30}
   l^{\pm} _{n} = \frac{c_{\pm}}{24} \left( \frac{r_{+} \mp r_{-}}{ l } \right) ^{2} \delta _{n,0}.
\end{equation}
The eigenvalues of the Virasoro generators $\hat{L} ^{\pm} _{n} $ are related to the energy $E$ and the angular momentum $J$ of the given black hole by the following formulae
\begin{equation}\label{7.31}
\begin{split}
   E =& l^{-1} ( l^{+} _{0} + l^{-} _{0} ) \\
=& \frac{1}{8 G} \left[ \left(\sigma + \frac{\alpha H}{\mu} + \frac{F}{m^{2}}\right) \frac{r_{+}^{2}+r_{-}^{2} }{l^{2}} + \frac{2 r_{+} r_{-}}{\mu l^{3}} \right] ,
\end{split}
\end{equation}
\begin{equation}\label{7.32}
\begin{split}
    J=& l^{-1} (l^{+} _{0} - l^{-} _{0}) \\
=&  \frac{1}{8 G} \left[ \left(\sigma + \frac{\alpha H}{\mu} + \frac{F}{m^{2}}\right) \frac{2 r_{+} r_{-}}{ l }  + \frac{r_{+}^{2}+r_{-}^{2} }{ \mu l^{2}} \right] .
\end{split}
\end{equation}
We can obtain the entropy of the BTZ black hole solution by the Cardy's formula \cite{61,62} (see also \cite{63} )
\begin{equation}\label{7.34}
\begin{split}
   S = &  2 \pi \sqrt{\frac{c_{+} l^{+} _{0} }{6}} + 2 \pi \sqrt{\frac{c_{-} l^{-} _{0} }{6}}\\
=& \frac{\pi}{2 G} \left[ \left(\sigma + \frac{\alpha H}{\mu} + \frac{F}{m^{2}}\right) r_{+} + \frac{r_{-}}{\mu l} \right] .
\end{split}
\end{equation}
 Since $\Omega _{H}=\frac{r_{-}}{l r_{+}}$ and $\kappa = \frac{r_{+}^{2}-r_{-}^{2}}{l^{2} r_{+}}$ are respectively the angular velocity and the surface gravity of horizon of BTZ black hole then one can check that equations \eqref{7.31}, \eqref{7.32} and \eqref{7.34} satisfy the first law of black hole mechanics.\\
Stationary black hole spacetime admits two Killing vectors $\partial_{t}$ and $\partial_{\phi}$. The conserved charges associated with the Killing vectors $\partial_{t}$ and $-\partial_{\phi}$ refer to mass, $M=Q(\partial_{t})$, and angular momentum, $j=Q(-\partial_{\phi})$, respectively. Also, entropy of black hole can be obtained by Eq.\eqref{21}.\\
Now we consider BTZ black hole \eqref{2.4.2} solution of EG. Since all the solutions of EG solve equations of motion of the GMMG then BTZ black hole will be a solution of the given model. Thus, using Eq.\eqref{17}, one can find energy and angular momentum of BTZ black hole as Eq.\eqref{7.31} and Eq.\eqref{7.32}, respectively. Using equations \eqref{22}, \eqref{37} and \eqref{38}, one can show that Eq.\eqref{21} in the GMMG reduces to
\begin{equation}\label{9.14}
  S =  \frac{1}{4G} \int_{r=r_{H}} \frac{d\phi}{\sqrt{g_{\phi \phi}}} \left[  \left(\sigma + \frac{\alpha H}{\mu}+ \frac{F}{m^{2}} \right) g _{\phi \phi} - \frac{1}{\mu} \Omega _{\phi \phi} \right]  ,
\end{equation}
where $r_{H}$ is the radius of horizon. By substituting Eq.\eqref{2.4.2} into the Eq.\eqref{9.14}, we find that the entropy of the BTZ black hole in the GMMG is as Eq.\eqref{7.34}.
\section{Further applications}
Some works have been done about the application of Eq.\eqref{17} and Eq.\eqref{21} in the context of GMMG and we want to list them here.\\
Asymptotically spacelike warped AdS$_{3}$ spacetimes are the solutions of GMMG \cite{33} and asymptotically admit $SL(2,\mathbb{R}) \times U(1)$ as an isometry group. Using the presented method, we have shown in \cite{33} that the algebra among the asymptotic conserved charges, in asymptotically spacelike warped AdS$_{3}$ spacetimes, is isomorphic to the semi-direct product of the Virasoro algebra with $U(1)$ current algebra. Warped black holes \cite{40} are also solutions of GMMG, and one can use the warped Cardy formula \cite{44,64} to find entropy of these black holes \cite{33}. In the paper \cite{33}, we showed that the entropy of warped black holes obtained by Eq.\eqref{21} is the same as what one can calculated by warped Cardy formula. Also, we applied Eq.\eqref{17} to find energy and angular momentum of warped black holes in the GMMG model. \\
The black flower \cite{46} is an exact solution of GMMG model. It was shown that the algebra among the conserved charges of this spacetime is isomorphic to two $U(1)$ current algebras with the levels $ \pm \frac{c_{\pm}}{12} $ \cite{29,32}. Also, in Ref.\cite{32}, we have used Eq.\eqref{17} to find conserved charges of Ba\~nados geometries.
\section{Near horizon conserved charges}
Near horizon geometry of a 3D black hole which has been considered in \cite{42} satisfies GMMG equations of motion near to the horizon. In this case, we treat the near horizon conserved charges as Noether charges. We avoid using Eq.\eqref{17} to obtain the near horizon conserved charges because it provide asymptotic conserved charges (conserved charges of the whole spacetimes). To this end, we integrate of the off-shell Noether potential Eq.\eqref{10} over horizon $H$
\begin{equation}\label{33}
  Q_{N}(\xi)= -\frac{1}{8\pi G} \int_{H} K_{\xi}.
\end{equation}
The algebra among the near horizon conserved charges is a semi-direct product of the Witt algebra with an Abelian current algebra which contains a central extension term (see \cite{43} for details).
\section{Conclusion}
This article provides a derivation of the conserved charges and entropy formula for black hole solutions in gravity theories defined by a Chern-Simons gravitational action in 3D. It was pointed out in \cite{22} that the derivation of the classical Wald formula for entropy is problematic in the first order formalism using the spin connection. In \cite{22} a method was introduced, based on the so-called LL derivative, to overcome this difficulty. Here, we extended such method to the case of a (generalized) Chern-Simons model. We have obtained not only the entropy formula but also a formula for calculation of the conserved charges, with the new method and have explicitly computed them in some models. We have listed some applications of Eq.\eqref{17} and \eqref{21}. The obtained results are consistent with literature.
\section{Acknowledgments}
M. R. Setare  thanks Dr. A. Sorouri for his help in improvement the English of the text.
\appendix
\section{Definition of Lorentz-Lie derivative}
Dreibein 1-form transforms like
\begin{equation}\label{A1}
  \tilde{e}^{a}= \Lambda^{a}_{\hspace{1.5 mm} b} e^{b}
\end{equation}
under Lorentz gauge transformations. By taking ordinary Lie derivative from Eq.\eqref{A1}, we have
\begin{equation}\label{A2}
        \pounds_{\xi}\tilde{e}^{a}= \Lambda^{a}_{\hspace{1.5 mm} b} \pounds_{\xi} e^{b} + \pounds_{\xi}\Lambda^{a}_{\hspace{1.5 mm} b} e^{b}
\end{equation}
Because of the presence of second term in the R.H.S. of Eq.\eqref{A2}, the ordinary Lie derivative of dreibein is not covariant under Lorentz gauge transformations. To modify the ordinary Lie derivative so that it does transform covariantly, we shall introduce $\lambda^{\hspace{1 mm} a}_{\xi \hspace{1 mm} b}$ such that it transforms like
\begin{equation}\label{A3}
    \tilde{\lambda}^{\hspace{1 mm} a}_{\xi \hspace{1 mm} b} = \Lambda^{a}_{\hspace{1.5 mm} c} \lambda^{\hspace{1 mm} c}_{\xi \hspace{1 mm} d} \left(\Lambda^{-1}\right)^{d}_{\hspace{1.5 mm} b} + \Lambda^{a}_{\hspace{1.5 mm} c} \pounds_{\xi} \left(\Lambda^{-1}\right)^{c}_{\hspace{1.5 mm} b},
\end{equation}
under Lorentz gauge transformations. Therefore,
\begin{equation}\label{A4}
    \pounds_{\xi}\tilde{e}^{a} + \tilde{\lambda}^{a}_{\hspace{1.5 mm} b} \tilde{e}^{b}= \Lambda^{a}_{\hspace{1.5 mm} b} \left( \pounds_{\xi} e^{b}+ \lambda ^{b}_{\hspace{1.5 mm} c} e^{c}\right).
\end{equation}
In this way, we can define a modified version of Lie derivative which does transform covariantly under Lorentz gauge transformation. We refer to the modified version of Lie derivative as Lorentz-Lie derivative which is defined by
\begin{equation}\label{A5}
  \mathfrak{L}_{\xi} e^{a}=\pounds_{\xi} e^{a}+ \lambda ^{\hspace{1 mm} a}_{\xi \hspace{1 mm} b} e^{b}
\end{equation}
This definition can be generalized for an arbitrary Lorentz tensor-valued $p$-form. For example, consider Curvature 2-form $R^{a}_{\hspace{1.5 mm} b} =d \omega^{a}_{\hspace{1.5 mm} b} + \omega^{a}_{\hspace{1.5 mm} c} \wedge \omega^{c}_{\hspace{1.5 mm} b}$. The ordinary Lie derivative of $R^{a}_{\hspace{1.5 mm} b}$ does not transform covariantly, i.e.
\begin{equation}\label{A6}
\begin{split}
  \pounds_{\xi} \tilde{R}^{a}_{\hspace{1.5 mm} b}= &  \Lambda ^{a}_{\hspace{1.5 mm} c} \pounds_{\xi} R ^{c}_{\hspace{1.5 mm} d} (\Lambda^{-1})^{d}_{\hspace{1.5 mm} b} + \pounds_{\xi}\Lambda ^{a}_{\hspace{1.5 mm} c} R ^{c}_{\hspace{1.5 mm} d} (\Lambda^{-1})^{d}_{\hspace{1.5 mm} b}\\
&+ \Lambda ^{a}_{\hspace{1.5 mm} c} R ^{c}_{\hspace{1.5 mm} d} \pounds_{\xi} (\Lambda^{-1})^{d}_{\hspace{1.5 mm} b}.
\end{split}
\end{equation}
By using transformation law of $\lambda ^{\hspace{1 mm} a}_{\xi \hspace{1 mm} b}$ \eqref{A3}, one can check that $\mathfrak{L}_{\xi}R^{a}_{\hspace{1.5 mm} b}=\pounds_{\xi} R^{a}_{\hspace{1.5 mm} b} + \lambda ^{\hspace{1 mm} a}_{\xi \hspace{1 mm} c} R ^{c}_{\hspace{1.5 mm} b} - R ^{a}_{\hspace{1.5 mm} c} \lambda ^{\hspace{1 mm} c}_{\xi \hspace{1 mm} b} $ transforms like $ \mathfrak{L}_{\xi}R ^{a}_{\hspace{1.5 mm} b} \rightarrow \Lambda ^{a}_{\hspace{1.5 mm} c} \mathfrak{L}_{\xi}R ^{c}_{\hspace{1.5 mm} d} (\Lambda^{-1})^{d}_{\hspace{1.5 mm} b} $ which is the transformation law for a Lorentz 2-tensor-valued 2-form. Therefore we can define the Lorentz-Lie derivative (LL-derivative) of a Lorentz tensor-valued $p$-form as $\mathfrak{L}_{\xi} \mathcal{A}^{a \cdots}_{b \cdots}= \pounds_{\xi} \mathcal{A}^{a \cdots}_{b \cdots} + \lambda^{\hspace{1 mm} a}_{\xi \hspace{1 mm} c} \mathcal{A}^{c \cdots}_{b \cdots}+ \cdots - \lambda^{\hspace{1 mm} c}_{\xi \hspace{1 mm} b}\mathcal{A}^{a \cdots}_{c \cdots} - \cdots $. The ordinary Lie derivative is involved with generator of infinitesimal diffeomorphism $\xi$. Therefore we expect that LL-derivative must involve with generator of infinitesimal Lorentz gauge transformations $\lambda ^{\hspace{1 mm} a}_{\xi \hspace{1 mm} b}$ as well as $\xi$ (which can be justified by definition of total derivative of a Lorentz tensor-valued $p$-form).

\end{document}